\documentclass[12pt,preprint]{emulateapj}
\usepackage{apjfonts}

\newcommand{\mps}{m\,s$^{-1}$}

\shorttitle{Effects of Active Region Flows on Meridional Flows}
\shortauthors{\v{S}vanda et al.}

\begin{document}

\title{Effects of Solar Active Regions on Meridional Flows}

\author{Michal \v{S}vanda}
\affil{Astronomical Institute, Charles University, Prague, CZ-18200, Czech Republic}
\affil{Astronomical Institute (v.v.i.), Academy of Sciences, Observatory Ond\v{r}ejov, CZ-25165, Czech Republic}
\email{michal@astronomie.cz}

\and

\author{Alexander G. Kosovichev and Junwei Zhao}
\affil{W. W. Hansen Experimental Physics Laboratory, Stanford University, Stanford, CA-94305-4085, USA}
\email{sasha@quake.stanford.edu, junwei@quake.stanford.edu}

\begin{abstract}
The aim of this paper is to extend our previous study of the solar-cycle variations of the meridional flows and to investigate their latitudinal and longitudinal structure in the subphotospheric layer, especially their variations in magnetic regions. Helioseismology observations indicate that mass flows around active regions are dominated by inflows into those regions. On average, those local flows are more important around leading magnetic polarities of active regions than around the following polarities, and depend on the evolutionary stage of particular active regions. We present a statistical study based on MDI/SOHO observations of 1996--2002 and show that this effect explains a significant part of the cyclic change of meridional flows in near-equatorial regions, but not at higher latitudes. A different mechanism driving solar-cycle variations of the meridional flow probably operates.
\end{abstract}


\section{Introduction}
Meridional flows are one of the integral properties of the solar convective envelope dynamics. They are essential for the magnetic flux transport from the activity belt towards the solar poles and therefore play a significant role in the polar field reversals \citep[e.g.][]{1989Sci...245..712W}. The average meridional flow profile (MFP) derived by local helioseismology is often used as an observational input in models describing the solar dynamo and magnetic flux transportation \citep[e.g.][]{2006ApJ...649..498D}. As we pointed out in our previous study \citep{2007ApJ...670L..69S}, it seemed that the flows around large active regions affect the mean MFP obtained as a longitudinal average of the North-South component of the flow velocity map. In this paper we investigate this effect in more detail and show that it may explain most of the observed variations of the meridional flow speed in low-latitude region, but not at high latitudes.

\section{Data and Analysis}
Local helioseismology methods provide an unique information about the subsurface dynamics of the Sun \citep[e.g.][]{2004SoPh..220..371H}. In particular, the time-distance helioseismology \citep{duvall93} is a tool that allows to use inversions of solar oscillation observations to infer the structure and topology of subsurface flows \citep{1996ApJ...461L..55K}. Solar acoustic waves ($p$~modes) are believed to be excited in the upper convection zone.  They travel between various points on the surface through the interior and are perturbed by mass flows along the path of propagation. The mass flow velocities in the interior are inferred from the differences of travel times from the central point to the surrounding annuli and the travel times from the surrounding annuli to the central point. 

The travel-time inversions are applied to full-resolution full-disk 1-min-cadence data observed by Michelson Doppler Imager \citep[MDI;][]{1995SoPh..162..129S} on-board the Solar and Heliospheric Observatory (SoHO). These data are available for only approximately two months a year uninterruptedly. From those data seven non-consecutive synoptic flow maps covering seven solar rotations during the period of 1996--2002 were constructed \citep{2004ApJ...603..776Z}. The data from the declining phase of the solar cycle were not used because they contain some probably instrumental issues and we decided not to use them for this particular study. For study of the relationship between the flows and magnetic fields synoptic maps of the line-of-sight component of magnetic field were constructed using the same algorithm based on MDI magnetograms.

In seven processed Carrington rotations we selected 92 magnetic active regions to investigate the influence of the flows in and around them to the total longitudinally averaged meridional flows by applying masks to isolated local flow patterns. For a proper masking, we need to determine the area of influence of active regions. We started with a rectangular box around each selected region, in which the magnitude of the line-of-sight component of magnetic field is above 25~Gauss. In this box we calculated a characteristic perturbation $\delta v$, defined as a mean of the absolute differences of meridional speed with respect to MFP calculated separately of each Carrington rotation, when the box around particular active region was masked. Then we expanded the box around this region while calculating the deviations $\delta v$ in this box. The expanded box, where $\delta v$ reaches its first maximum, we considered as an area of influence of particular active region. Other maxima are attributed to the presence of other magnetic region, which appeared in the expanded box. The regions of influence around each magnetic area were masked in the synoptic map. The size of the region of influence was on average twice as large as the initial 25-Gauss box. 

\section{Results}
\begin{figure}
\epsscale{1.}
\plotone{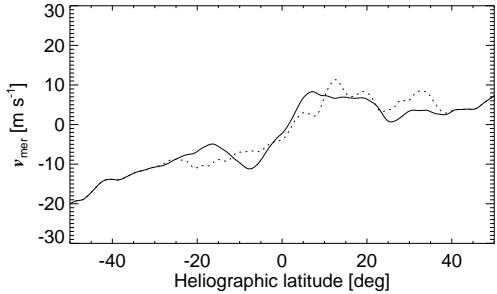}
\caption{The longitudinally averaged meridional component of the time-distance helioseismology flow map at 3--4.5~Mm depth (solid line) and  the same quantity when active regions are masked (dotted line) for Carrington rotation No.~1975 (April 2001). Error-bars have size of $\sim$2~\mps.\label{fig1}}
\end{figure}

\paragraph{Variations caused by active regions}
The new analysis confirmed the results of our previous study \citep {2007ApJ...670L..69S}. The local flows around strong active regions affect the mean meridional flow obtained as a longitudinal average of the North-South component of the synoptic flow map. On average, the large-scale inflow pattern in the activity belt prevails, which is in agreement with findings by \cite{2003ApJ...585..553B}, \cite{2004ApJ...603..776Z}, \cite{2004SoPh..220..371H}, and \cite{2006AdSpR..38..845K}. For example, in Fig.~\ref{fig1} one can clearly see that MFP calculated excluding active regions (henceforth called as the reference meridional flow profile -- RMFP) varies less with latitude than when the flows associated with active regions are included. The deviations between the profiles with and without the active region flows are up to 5~\mps. On average, the positive deviations in the areas of influence prevail in the Northern hemisphere (by 0.08~\mps), while the negative ones prevail in the Southern hemisphere (by 0.10~\mps). This may be interpreted that the inflows into active regions are stronger on their equatorial side than on poleward one, in both hemispheres. Such small differences are not statistically significant; thus to confirm this, a detailed study is needed. An analytical model of torsional oscillations by \cite{2003SoPh..213....1S} predicted such inflows into the activity belt with characteristic amplitude of 6~\mps. Similar values were predicted also by numerical model by \cite{2006ApJ...647..662R}. Both models assumed a mechanism of thermal forcing for the low-latitude torsional oscillations branches. The inflow in the activity belt was a side-effect of these models, but it resembles the observed inflow in active regions.

\begin{figure*}
\epsscale{1.}
\plottwo{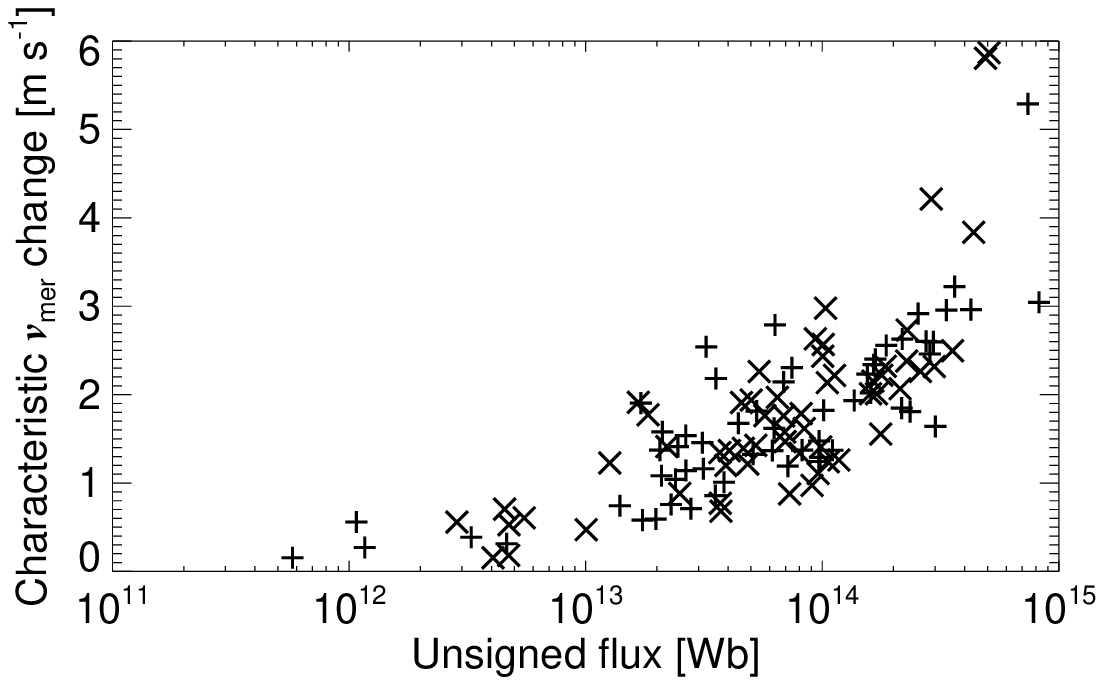}{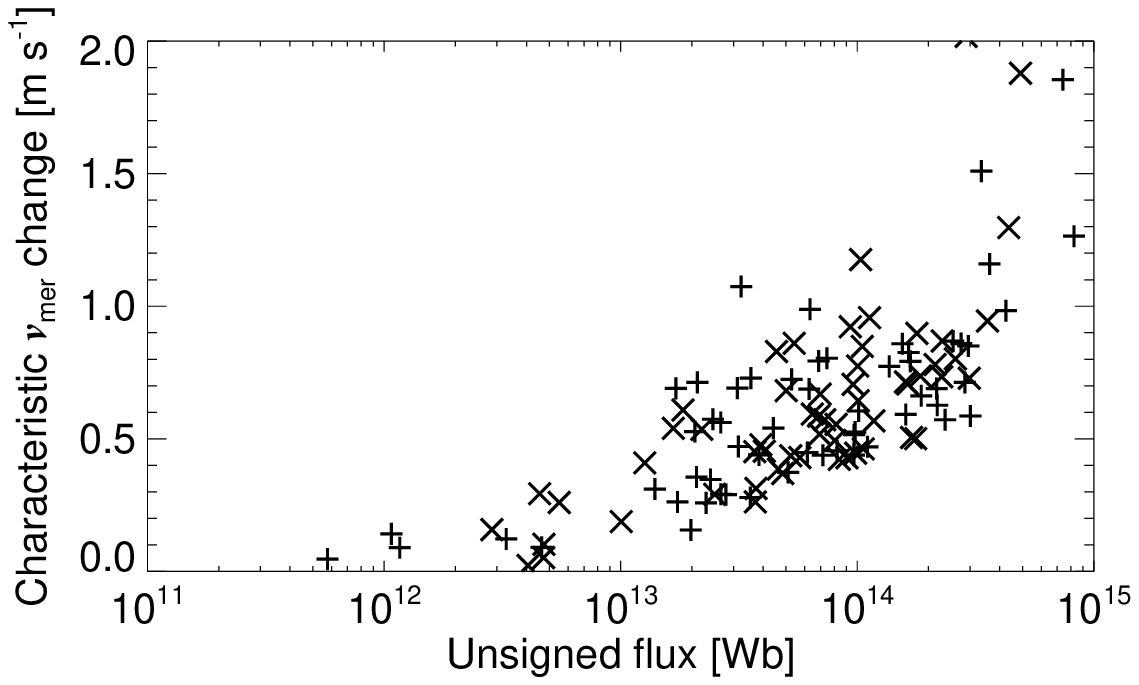}
\caption{A dependence of the characteristic variations of the meridional flow on the total unsigned flux localized in active regions for the flows at the depth of 0--3~Mm (left) and 3--4.5~Mm (right). The dependence in both depth intervals is very similar, but the magnitude of deviations is smaller at greater depth. The '+' signs are for areas in the Northern hemisphere, '$\times$' sings for the Southern hemisphere ones.\label{fig2}}
\end{figure*}

We investigated a possible dependence of the characteristic perturbation of the flows at different depths in individual active regions on the amount of magnetic flux present in their areas. As a measure of the characteristic perturbation $\delta v_{\rm mer}$ we used a mean of the absolute differences between the measured meridional component of the flow field and RMFP at given latitude calculated over the region of interest.

\begin{figure*}
\epsscale{0.9}
\plotone{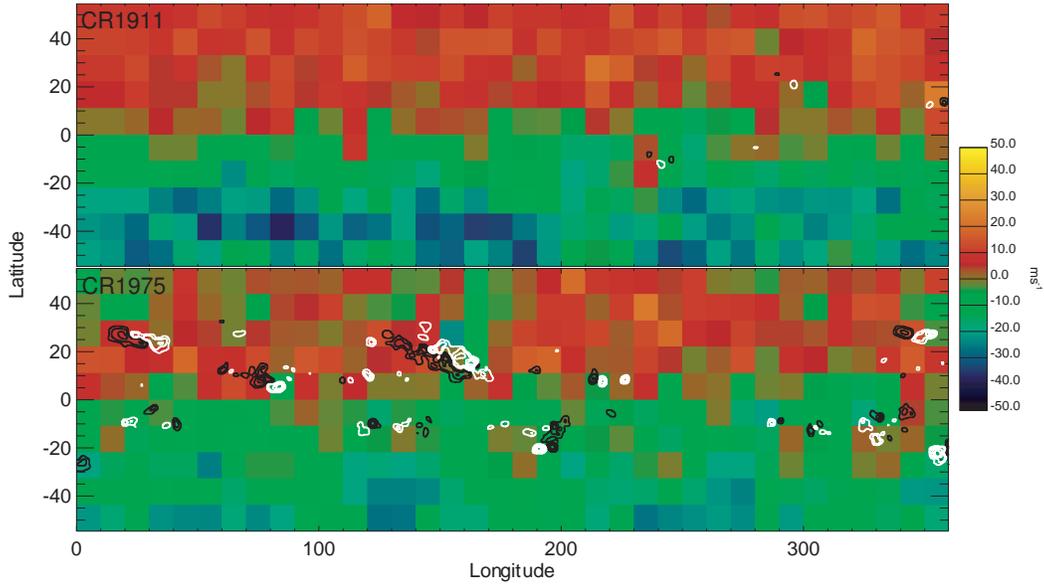}
\caption{Examples of synoptic maps of the meridional (North-South) component at the depth of 3--4.5~Mm smoothed and binned by 10 heliospheric degrees. The positive meridional speed is Northward. The magnetic active regions are displayed with contours: white contours show the positive polarity, the black ones show the negative polarity. These maps demonstrate a longitudinal structure of meridional flows close to minimum (CR~1911) and maximum (CR~1975) of solar activity. The errors of individual flow velocity measurements are $\sim$2~\mps{}  \citep{2004ApJ...603..776Z}.\label{fig3}}
\end{figure*}

The dependence of $\delta v_{\rm mer}$ on the total unsigned flux for all 92 active regions is shown in Fig.~\ref{fig2}. Other characteristics, such as a geometrical average of deviations over the whole active region, display a similar dependence. The perturbations become less significant with the increase of depth in the convection zone. Fig.~\ref{fig2} indicates that the active regions containing more magnetic flux cause more significant deviations from RMFP than those containing less flux.

Examples of synoptic maps of the meridional flow component for Carrington rotations CR~1911 (July 1996) and CR~1975 (April 2001) are shown in Fig.~\ref{fig3}. These maps contain a wide range of local motions of various scales, but their structure is far from the idealized single-cell picture. In the map of CR~1975 one can see very large and complex active regions with converging flows (inflows) prevailing in the leading parts of the active regions, while the trailing parts do not seem to differ in topology of flows from the quiet Sun regime. This effect was already suggested by \cite{2006IAUS..233..365K}. In those examples we see that converging flows extend from the active region far in the poleward direction and form a surge-shaped structure. 

\begin{figure}[b]
\epsscale{1}
\plotone{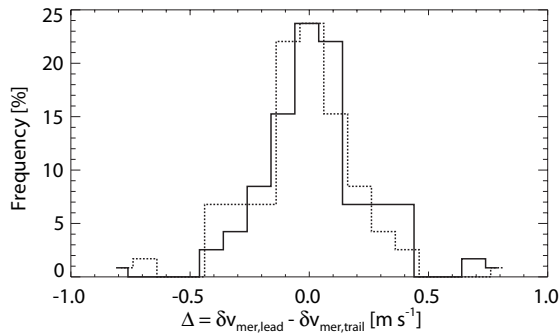}
\caption{Histogram of the characteristic deviations of meridional components from the RMFP, showing the difference between the leading and trailing polarities of active regions in our sample. The histogram reflected around $\Delta 0$ is show in dotted line to enhance the visibility of the asymmetry. \label{fig4}}
\end{figure}

\paragraph{Asymmetry between the flows around leading and following polarities of active regions}
The question is whether this asymmetry between the flow around the leading and following polarities is common for all active regions. To study this issue we divide the box around each active region in two parts separating the regions of the leading and following polarities. As a measure of the flow deviations in leading and trailing polarities we use the difference $\Delta$ between $\delta v_{\rm mer}$ in both polarities of each active region, i.e.
\begin{equation}
\Delta = \delta v_{\rm mer}({\rm leading})-\delta v_{\rm mer}({\rm trailing})\ .
\end{equation}
We do not find a clear correlation of $\Delta$ with any characteristic of magnetic field of the active regions. However, a histogram of $\Delta$ (Fig.~\ref{fig4}) shows that the leading and following polarities differ quite often as the positive part contains more cases than the negative one.

The distribution of $\Delta$ seems to be bimodal. The second peak (around 0.4~\mps) suggests an existence of group of active regions, where inflows in their leading polarities are significantly stronger than in trailing ones. Some examples of active regions, for which the flows have the behavior described above, can be seen in Fig.~\ref{fig3}. There is no typical characteristic or feature among the active regions responsible for the second peak in the distribution of $\Delta$ in our sample, except for the tilt angle, calculated from positions of the centres of mass of both polarities. The tilt angle is on average larger (38 degrees vs. 19 degrees) for the active regions that form the second peak in the distribution of $\Delta$. However, those numbers are not convincing due to the small sample. Perhaps, this effect can be related to the age of active regions. As pointed out, e.g. by \cite{1994SoPh..149...23H}, the tilt angle of bipolar sunspot groups on average may increase during their evolution.

Our visual inspection shows that most of 11 active regions making the second peak of the histogram have a trailing polarity dispersed probably by diffusion, while the leading one seems to be more compact. A quick look on SolarMonitor.org and Mt.~Wilson drawings archive showed that these active regions are old and highly evolved with complex topology. A detailed investigation of this effect requires a larger dataset of helioseismic and magnetic observations of active regions. 

\begin{figure}
\epsscale{1.05}
\plotone{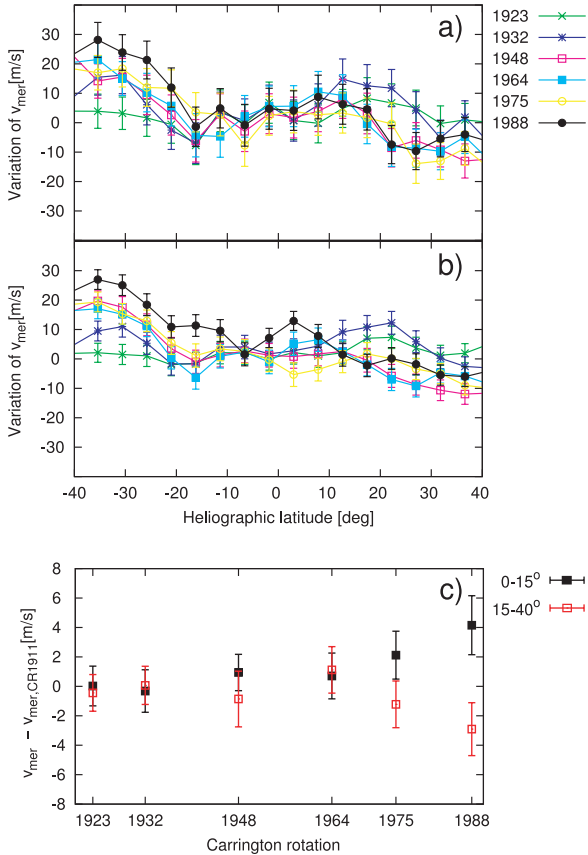}
\caption{\emph{a)} -- Solar-cycle variations of MFP averaged over depths of 0--9~Mm after the subtraction of MFP in CR1911. \emph{b)} -- The same with active regions masked in the maps. \emph{c)} -- The difference of a) and b) averaged over two latitudinal belts to demonstrate the effect of the active regions on longitudinally averaged MFP at various phases of the magnetic activity (the negative values resemble equatorward flow).\label{fig5}}
\end{figure}

\paragraph{Latitudinal variations of the meridional flows}
In Fig.~\ref{fig5}, the evolution of the RMFP averaged over depths of 0--9~Mm with solar-cycle is given, from which the profile in minimum of solar activity (CR1911) was subtracted. We basically see that the meridional flow slows down at higher latitudes with the increasing magnetic activity, and that variations are larger when the magnetic regions are included in the synoptic flow maps. The differences between corresponding curves in Figs.~\ref{fig5}a and ~\ref{fig5}b are statistically significant. In Fig.~\ref{fig5}c the variations caused by the magnetic regions averaged over two latitudinal belts in both hemispheres are given. The opposite sign of the meridional flow in both hemisphere was taken into account. We see that the deviations from RMFP caused by the active regions increase with activity cycle, which is consistent with results described in previous sections. 

A large amount of cycle-related variation can be explained by flows around the active regions in near-equatorial belt. If we quantify this effect calculating value $\vartheta=(v_{\rm mer}-v_{\rm mer,ref})/(v_{\rm mer}-v_{\rm mer,CR1911})$, where $v_{\rm mer}$ is MFP including flows around the active regions, $v_{\rm mer,ref}$ the one with active regions masked, and $v_{\rm mer,CR1911}$ is MFP in minimum of solar activity, we may estimate the importance of the flows around the active regions on the cycle-related variations of MFP. For the equatorial region (latitudes of 0 to 15 degrees) we find $\vartheta=(0.65\pm0.21)$ in the Southern and $\vartheta=(0.34\pm0.26)$ in the Northern hemisphere, and for higher latitudes (15 to 40 degrees) we find $\vartheta=(0.20\pm0.15)$ for the Southern and $\vartheta=(0.37\pm0.29)$ for the Northern hemisphere. The North-South asymmetry is statistically significant, the effect is stronger in the Southern hemisphere, where also the surface magnetic activity is stronger than on the Northern hemisphere. On average, the effect of the flows around the active regions is not sufficient to explain cycle-related changes of MFP, especially at higher latitudes, where remaining cycle-related variations may be explained by formation of persistent high-latitude counter-cells providing the mass inflow on poleward side of activity belt. The peak speed in such counter-cell could be estimated by $\sim$10~\mps{} in the Northern hemisphere and $\sim$25~\mps{} in the Southern hemisphere. Perhaps, this may be due to the existence of deep magnetic field, which could also depict North-South asymmetry. We think that these results represent an interesting challenge for theories and numerical simulations of the solar meridional circulation.

\paragraph{Longitudinal variations of RMFP}
The question whether or not the MFP without active regions is the same at different longitudes naturally arises. Unfortunately, we cannot provide a clear answer based on the available data. As we can see in Fig.~\ref{fig3}, the synoptic maps of the large-scale North-South flows component show a lot of variations, which are not persistent for the whole synoptic map. These local disturbances of the meridional component of the flow have greater magnitude than the mean flow. The magnitude of the disturbances is decreasing with depth. One needs to keep in mind that in general, it is impossible to distinguish between the temporal evolution of the meridional flow and its longitudinal structure in synoptic map, which is composed of daily strips around the central meridian. However, the assumption that persistent component of meridional flow evolves slowly with characteristic time of several days seems reasonable, which justifies our approach. It was described in numerical simulations \citep[e.g.][]{2008ApJ...673..557M} that meridional flow has a strong fluctuating component, magnitude of which is larger than magnitude of the mean temporally averaged meridional flow. Some fluctuations surviving for many hours are also detected in our data and the mean poleward meridional flow appears only after averaging. Perhaps, similar fluctuations happen in the real Sun as well.

\section{Conclusions}

From our study it becomes clear that the large-scale mass flows around active regions significantly influence the mean meridional flow. This is an important conclusion for the dynamo and solar cycle models, which use the longitudinally averaged meridional flow from local helioseismology as a measure of the mean meridional flow. Profiles calculated simply by taking a longitudinal average do not represent the mean meridional circulation in and between activity belts. For particular active region, the deviations caused by the circulation flows around them depend on the total magnetic flux in the area. On average, the local circulation flows are represented by inflows in the activity belt, as predicted by \cite{2003SoPh..213....1S} and observed by i.e. \cite{2003ApJ...585..553B} or \cite{2004ApJ...603..776Z}.

Moreover, it seems that, on average, the local flows are more important in the areas of leading magnetic polarities of active regions. There is a group of active regions in our sample, where local flows are significantly more important in the leading polarities. This effect may be related to the evolution of active regions and requires a detailed study based on extended dataset describing three-dimensional structure of flows in and around active regions in various stages of their evolution. Those deviations explain a significant part of the cyclic change of the meridional profile shape in the equatorial region, but not at higher latitudes. A different mechanism must operate to explain all cyclic variations. Using the currently available dataset we find an evidence that the latitudinal variations of the meridional flow at high latitudes may be related to subsurface magnetic activity, but at present state this is inconclusive. Another interesting problem is how the asymmetry in the strength of the local inflows around leading and following polarities of active regions affects the magnetic flux transport to the polar regions and Sun's polarity reversals. As we pointed our in
\cite{2007ApJ...670L..69S}, this phenonenon is not important for the meridional flux transport averaged over a few Carrington rotations. We expect that in detail this conclusion may be changed. This problem can be studied when more local helioseismology data will become available from Solar Dynamics Observatory.
 
\acknowledgments
M.\v{S} acknowledges a kind support and hospitality of SOI group at Stanford University and also support of GA~AV\v{C}R (grant IAA30030808), and ESA-PECS (grant 98030). A\'U~UK works on Research Program MSM0021620860 (M\v{S}MT \v{C}R).


\begin{thebibliography}{}
\bibitem[Basu \& Antia(2003)]{2003ApJ...585..553B} Basu, S., \& Antia, H.~M.\ 2003, \apj, 585, 553 
\bibitem[Dikpati \& Gilman(2006)]{2006ApJ...649..498D} Dikpati, M., \& Gilman, P.~A.\ 2006, \apj, 649, 498
\bibitem[Duvall et al.(1993)]{duvall93} Duvall, T.L., Jr., Jefferies, S.M., Harvey, J.W., Pomerantz, M.A., 1993, Nature, 362, 430
\bibitem[Haber et al.(2004)]{2004SoPh..220..371H} Haber, D.~A., Hindman, B.~W., Toomre, J., \& Thompson, M.~J.\ 2004, \solphys, 220, 371 
\bibitem[Howard(1994)]{1994SoPh..149...23H} Howard, R.~F.\ 1994, \solphys, 149, 23 
\bibitem[Komm et al.(2006)]{2006AdSpR..38..845K} Komm, R., Howe, R., \& Hill, F.\ 2006, Advances in Space Research, 38, 845 
\bibitem[Kosovichev(1996)]{1996ApJ...461L..55K} Kosovichev, A.~G.\ 1996, \apjl, 461, L55 
\bibitem[Kosovichev \& Duvall(2006)]{2006IAUS..233..365K} Kosovichev, A.~G., \& Duvall, T.~L.\ 2006, Solar Activity and its Magnetic Origin, 233, 365 
\bibitem[Miesch et al.(2008)]{2008ApJ...673..557M} Miesch, M.~S., Brun, A.~S., DeRosa, M.~L., \& Toomre, J.\ 2008, \apj, 673, 557 
\bibitem[Rempel(2006)]{2006ApJ...647..662R} Rempel, M.\ 2006, \apj, 647, 662 
\bibitem[Scherrer et al.(1995)]{1995SoPh..162..129S} Scherrer, P.~H., et al.\ 1995, \solphys, 162, 129 
\bibitem[Spruit(2003)]{2003SoPh..213....1S} Spruit, H.~C.\ 2003, \solphys, 213, 1 
\bibitem[{\v S}vanda et al.(2007)]{2007ApJ...670L..69S} {\v S}vanda, M., Kosovichev, A.~G., \& Zhao, J.\ 2007, \apjl, 670, L69 
\bibitem[Wang et al.(1989)]{1989Sci...245..712W} Wang, Y.-M., Nash, A.~G., \& Sheeley, N.~R., Jr.\ 1989, Science, 245, 712 
\bibitem[Zhao \& Kosovichev(2004)]{2004ApJ...603..776Z} Zhao, J., \& Kosovichev, A.~G.\ 2004, \apj, 603, 776 
\end{thebibliography}
\end{document}